\documentclass[12pt]{article}
\usepackage[T1]{fontenc}

\makeatletter

\newcommand{\LyX}{L\kern-.1667em\lower.25em\hbox{Y}\kern-.125emX\@}

\usepackage{amscd,amssymb}
\usepackage{epsfig}
\makeatother

\begin{document}

\title{IS SUPERSYMMETRIC QUANTUM MECHANICS COMPATIBLE WITH DUALITY ?}

\maketitle
{\large \hfill{}M. CAPDEQUI PEYRANÈRE\hfill{} }\\
{\large \par}

{\par\centering \textit{\small Laboratoire de Physique Mathématique et Théorique} \par}

{\par\centering {\small CNRS - U}\textit{\small MR 5825 }\small \par}

{\par\centering \textit{Université de Montpellier II, 34095 Montpellier Cedex.5
(France)} \par}

\begin{abstract}
Supersymmetry applied to quantum mechanics has given new insights in various
topics of theoretical physics like analytically solvable potentials, WKB approximation
or KdV solitons. Duality plays a central role in many supersymmetric theories
such as Yang-Mills theories or strings models. We investigate the possible existence
of some duality within supersymmetric quantum mechanics. 
\end{abstract}
\vspace{4.5cm}
\hspace{1.5cm} PM/99-43. To be published in Modern Physics Letters A\\

\newpage

\textbf{Introduction}\\

\renewcommand{\theequation}{\thesection.\arabic {equation}}

Supersymmetry is an exciting idea to relate bosons and fermions. Many particle
physicists think that supersymmetry could be found in Nature, not too far from the
electroweak scale, typically the TeV scale. From a theoretical point of view,
supersymmetry naturally explains, for example, the large ratio between the electroweak
scale and the other higher physical scales as the Planck or the GUT scales.
Moreover, supersymmetry favours the unification for the three fundamental coupling
constants responsible for the electromagnetic, the weak and the strong interactions.
With the present data, such an unification is not possible within the Standard
Model but becomes probable in its supersymmetric version like the MSSM (Minimal
Supersymmetric Standard Model). Supersymmetry can also enrich quantum mechanics
and gives the so called supersymmetric quantum mechanics (SQM) which is a nice
framework to test new concepts in Physics. SQM is famous due to the pioneering papers 
of Witten on supersymmetry breaking{[}1{]. SQM also gives  new understandings of the analytical
solvability of certain potentials and of the (super)BKW approximation which
turns out to be exact for a large class of potentials, or allows connections between
isospectral hamiltonians and the multi-soliton solution of the Korteweg-De Vries
non linear equation. Much more about SQM can be found in {[}2,3{]} and references therein. 

Duality is a very fascinating concept whose original aim was to connect
two different regimes of the same dynamics. The electric-magnetic duality is
the most popular example where a weakly coupled version of a theory is exchanged
for a strongly coupled formulation. The concept of duality has been extended
and means an equivalence between two formulations of a model or between two
different models. Such a case is the duality between the sine - Gordon model
and the massive Thirring model {[}4{]}, connecting a two-dimensional solitonic
object and a two-dimensional fermionic field. That means that there is no guarantee
at all that the connection is simple. 

Supersymmetry goes well with duality. As a matter of fact, some problems in
duality invariant gauge theories disappear when supersymmetry comes in. This
is exactely what happens in {[}N=4{]} Supersymmetric Yang-Mills theory {[}5{]}.
Supersymmetry is also an essential ingredient to study duality in string theories
{[}6{]}. In this paper we will elaborate on the following simple idea: is
it possible to find a strong/weak coupling duality in the framework of supersymmetric
quantum mechanics ? However clear the question may be, the definition remains vague.
Thus we will discuss a bit more this point later, in the first section of
this paper. In the second section we will develop duality in SQM with a precise definition.
After the conclusion, we will work out in an appendix two examples to illustrate
 section II.\\

\renewcommand{\theequation}{I.\arabic{equation}}

\textbf{Section I}\\

The idea we wish to explore is the mixing of duality and supersymmetric quantum
mechanics. What does this mean? To be more precise, we will try to find duality
in SQM and not to link by duality SQM to some other theory, and more exactly to
stay in a model of SQM, and not to map a model of SQM with a strong coupling
in another model of SQM with a weak coupling. 

Any model of SQM can be written in a superspace formalism like a field theory.
Here we will use the hamiltonian formulation where the physics of the model
lies in an hamiltonian \( H \) which is a non negative operator. One has \( H=\left\{ Q,Q^{*}\right\}  \)
where \( Q \) and \( Q^{*} \), the ``supercharges'', can be seen as operators
transforming \char`\"{}bosons\char`\"{} into \char`\"{}fermions\char`\"{} and
vice versa. Also any SQM hamiltonian can be cast in a matrix form as follows
:

\[
H=\left| \begin{array}{cc}
H_{-} & 0\\
0 & H_{+}
\end{array}\right| \]

These two hamiltonians \( H_{-} \) and \( H_{+} \) are connected by supersymmetry
which in turn gives a relation between the potential \( V_{-} \) and \( V_{+} \)
via the so called superpotential \( W \) {[}2,3{]}. The question is to know
whether one of them can be strongly coupled and the other one weakly coupled. 

Let us denote by \( x \) the (one dimensional) space variable and by g a real positive
coupling constant. One has :

\[
V_{-}(x,g)=W^{2}(x,g)-W'(x,g)\]
 and

\[
V_{+}(x,g)=W^{2}(x,g)+W'(x,g),\]
 where \( W'=dW/dx. \)

Naively, if \( V_{-} \) is zero, the coupling could be said \char`\"{}weak\char`\"{}
; then \( V_{+} \) will be different from zero and the coupling could be said
\char`\"{}strong\char`\"{}. The solution is \( W=-1/(a+x) \), where a is an
arbitrary constant, and \( V_{+} \) is obviously \( 2/(a+x)^{2} \). As a consequence
of this rough example, it seems possible to see a path through the solution
but perhaps not really interesting. Indeed, in this example, there are some
drawbacks :

- \( g \) does not appear at all ; 

- the potential \( V_{+} \) is a singular potential on the line ; 

- the Hamiltonian \( H_{-} \) is just a second derivative, without any dynamical content. \\
One can be a bit less naive in modifying our guess : is it possible to find
a couple of supersymmetric potentials such as :



\begin{equation}
V_{-}(x,g)=g\, f(x,g)\ \ \ \mbox{and}\ \ \  
V_{+}(x,g)=f(x,g)/g\ ?
\end{equation}

In such a way it seems that one forces, if \( g\sim 0 \), a weak coupling for
\( V_{-} \) and a strong coupling for \( V_{+} \). In solving the system (I.1),
one successively gets the superpotential :

\[
W(x,g)=\frac{-1}{s(a+x)}\, \, \, ,\]
 and the potentials:

\[
V_{-}(x,g)=\frac{(1-s)}{s^{2}(a+x)^{2}}\, \, ,\]
 and

\[
V_{+}(x,g)=\frac{(1+s)}{s^{2}(a+x)^{2}}\, \, ,\]
 where \( s=\left( 1-g^{2}\right) /\left( 1+g^{2}\right)  \) and a is an arbitrary
constant. 

 Although the potentials are still singular
on a line, this result is quite interesting. Indeed, one obtains the following
 identity :

\[
V_{-}(x,1/g)=V_{+}(x,g)\]
since s is changed in \( -s \) when one changes \( g \) in \( 1/g \).
This
is a rather good feature. Now both potentials depend only on s and not on \( g \),
and this is a rather bad feature, because when g runs from zero to infinity,
\( s \) goes between \( -1 \) and \( +1 \) and the strength of the couplings
is somewhat washed off.

 This is a general consequence of our proposal to search for duality within the same SQM model. Actually, assuming there
is only one coupling constant \( g(\neq 1) \) in the model, the duality we
look for can be written as follows :

\begin{equation}
\label{I.2}
H_{-}(x,1/g)=H_{+}(x,g)
\end{equation}

That means that a SQM model determined by the two hamiltonians \( (H_{-}(x,1/g),\, H_{+}(x,1/g)) \),
connected by susy thanks to the superpotential \( W(x,1/g) \) in a regime
of strong coupling (if \( g\sim 0 \)), is exchanged in the SQM model determined
by \( (H_{+}(x,g),\, H_{-}(x,g)) \), which works in a weak coupling regime.
The formula I.2 is just equivalent to \( V_{-}(x,1/g)=V_{+}(x,g) \), or :

\[
W^{2}(x,1/g)-W'(x,1/g)=W^{2}(x,g)+W'(x,g).\]
 One can re-write this formula as :

\begin{equation}
\label{I.3}
W^{2}(x,1/g)-W^{2}(x,g)=W'(x,1/g)+W'(x,g)\, .
\end{equation}
 When one changes \( g \) in \( 1/g \) in this formula, the left hand side
gets the opposite value while the right hand side is invariant. Hence both sides vanish and the unique superpotential solution 
of I.2  is such that: 

\begin{equation}
\label{I.4}
W(x,1/g)=-W(x,g)\, .
\end{equation}
 So any function of \( x \) and \( g \) verifying I.4 is an answer to our
question. If one looks for such functions, the simplest form is \( W(x,g)=a(x)b(g) \),
with the condition

\begin{equation}
\label{I.5}
b(1/g)=-b(g)\, .
\end{equation}

The general solution of I.5 is \(b(g) = B(Log(g))\)  where \(B\) is any odd function, but one can find rather 
simple solutions without logarithm. For example, \( b(g)=g-1/g \) is a solution but not
 suitable for a perturbative point a view (as \(Log(g)\)) since it has no expansion
neither in \( g=0 \) nor in \( g= \) infinity. Better solutions are \( b(g)=(1-g)/(1+g) \),
which is \(tanh(-(Log(g))/2)\), or \( b(g)=arctan(g) -\pi /4 \). 

Anyway, the result (I.5) induces two remarks : 

- the true coupling constant is \( b(g) \) and not \( g \). 

- if the true coupling constant \( b(g) \) belongs to the intervalle \( \left[ c,d\right]  \),
then \( b(1/g) \) belongs to the intervalle \( \left[ -d,-c\right]  \). 

The consequence is evident: it is impossible to talk about strong/weak duality
in such a simple case. Of course, \( W(x,g)=a(x)\, b(g) \) is far from being the
most general solution of (I.4) but one cannot find other solutions which escape the
previous consequence, because for a fixed \( x \) the relation I.4 between
\( W(x,g) \) and \( W(x,1/g) \) falls in a formula similar to I.5. Thus it
is impossible to find a strong/weak duality in a same model of SQM, but one
can search for a class of models where the couple \( (H_{-},H_{+}) \) is changed
in the couple \( (H_{+},H_{-}) \), depending on the sign of the (true) coupling
constant value. In these cases, the duality would mean a correspondence between
two different regimes of the same SQM model.\\

\textbf{Section II }\\

\renewcommand{\theequation}{II.\arabic{equation}}

\setcounter{equation}{0}

In order to naturally introduce the class of models that we look for, we will
continue to follow for a while the ingenuous path presented in the previous
section. Further, we note that, for any quantum mechanical hamiltonian, there
are several supersymmetric extensions. For example, starting from the harmonic
oscillator hamiltonian \( H_{0} \) ( in convenient units : \( 2m=\hbar =1 \)
) :

\[
H_{0}=-d^{2}/dx^{2}+\omega ^{2}x^{2},\]
 one has

\[
H_{-}=-d^{2}/dx^{2}+\omega ^{2}x^{2}-\omega ,\]
 and

\[
H_{+}=-d^{2}/dx^{2}+\omega ^{2}x^{2}+\omega ,\]
 when the superpotential is \( W=\omega x \) . But it is possible to obtain
other superpotentials and hamiltonians \( H_{-} \) and \( H_{+} \) (see for
instance the second example of the Appendix ). So we will assume that we know a \char`\"{}free\char`\"{}
hamiltonian \( H_{0}=-d^{2}/dx^{2}+k(x) \), where \( k(x) \) is a fonction
of \( x \) and of some other parameters, and we add extra terms to simulate
the weak and strong couplings as we did in I.1. We write :

\[
H_{-}(x,g)=-d^{2}/dx^{2}+k(x)+g\, f(x,g)\]
 and

\begin{equation}
\label{II.1}
H_{+}(x,g)=-d^{2}/dx^{2}+k(x)+f(x,g)/g\, .
\end{equation}
 In accordance with the idea of \( H_{0} \) as a free hamiltonian, we suppose also that \( k(x) \) is independent of \( g \) (\( g\neq 1 \)).
One could probably imagine other forms that II.1 but the main virtue
of our type is its solvability, without altering the physical meaning.
One could also slightly modify the proposal II.1 by some sign or by multiplying
\( f \) by \( g \) ; these changes are really minor, thus we will not consider
them any longer . 

In solving II.1 in terms of the superpotential, one obtains the following Riccati
equation :

\begin{equation}
\label{II.2}
W'(x,g)=s\left( W^{2}(x,g)-k(x)\right) ,
\end{equation}
 where again we get the parameter \( s=\left( 1-g^{2}\right) /\left( 1+g^{2}\right)  \).
We use the usual trick to linearize the Riccati equation II.2 ; introducing
a function \( y \) of \( x \) such as \( W=-y'/(s\, y) \), one readily gets
the following Sturm - Liouville equation : 

\begin{equation}
\label{II.3}
y''-s^{2}k(x)y=0\, .
\end{equation}
 Once a solution \( y \)  is found for this equation II.3, we have the potentials
solution of the problem (II.1) :

\[
V_{-}(x,g)=+sk(x)+(1-s)(y'/(sy))^{2},\]
 and 

\begin{equation}
\label{II.4}
V_{+}(x,g)=-sk(x)+(1+s)(y'/(sy))^{2}.
\end{equation}

 The supersymmetry is not broken if one can choose
the sign of \( s \) such that only one of the following sets of equations is
true :

\[
H_{-}(x,g)\, y^{(+1/s)}=0\, \, \textrm{and}\, \, H_{+}(x,g)\, y^{(+1/s)}\neq 0,\]
 or

\[
H_{+}(x,g)\, y^{(-1/s)}=0\, \, \textrm{and}\, \, H_{-}(x,g)\, y^{(-1/s)}\neq 0.\]

Furthermore to get discrete spectra for \( H_{-} \) and \( H_{+} \), one
needs that \( y^{(1/s)} \) or \( y^{(-1/s)} \) be normalizable \( \left( L^{2}(x,R)\right)  \),
depending on the signe of \( s \). Then the corresponding hamiltonians \( H_{-} \) and \( H_{+} \) will be isospectral (up to the lowest bound state).

Let us be more precise about the solution II.4. At first sight, it seems that
when we perform the following change : \( g\rightarrow 1/g \), then \( s\rightarrow -s \)
and \( V_{-}(x,g)\rightarrow V_{-}(x,1/g) \), which would be equal to \( V_{+}(x,g) \).
Actually, this is not true because it is necessary to take into account the
eventuality that the boundary conditions of the equation II.3 depend on \( g \).
In such a way, \( y \) would be \( g \) dependent and \( V_{-}(x,1/g) \)
different from \( V_{+}(x,g) \). So, in general, \( V_{-}(x,1/g)\neq V_{+}(x,g) \)
and it is not suitable for duality ; fortunately there are simple situations
where we get it. 

Assuming that \( k(x) \) is an even function of \( x \), then the second order
differential equation II.3 (without first order term) has two independent solutions
of given parity, one is even and the other is odd in the variable \( x \),
and both depend on \( s^{2} \). The odd one is zero at \( x=0 \), so leading possibly to singular potentials.
Thus it is highly preferable to use the even solution, and then the ratio \( y'/y \)
will be well defined and insensitive to a possible \( g \) dependence coming from the boundary
conditions. This is a case where we get the wished duality : \\

- first regime :

\[
0<s<1\, \, \, \, (0<g<1)\]

\[
H_{-}(x,s)=-\frac{d^{2}}{dx^{2}}+V_{-}(x,s)\]

\[
V_{-}(x,s)=+sk(x)+(1-s)(y'/(sy))^{2},\]

\[
H_{+}(x,s)=-\frac{d^{2}}{dx^{2}}+V_{+}(x,s)\]

\[
V_{+}(x,s)=-sk(x)+(1+s)(y'/(sy))^{2},\]
 and (for example) \( H_{-}(x,s)\, y^{(+1/s)}=0 \) with \( H_{+}(x,s)\, y^{(+1/s)}\neq 0 \)
and \( y^{(+1/s)} \) is normalizable. In this regime, the ground state energy
of \( H_{-} \) is zero and the other energy eigenvalues of both \( H_{-} \)
and \( H_{+} \) are paired.\\

- second regime :

\[
-1<s<0\, \, \, \, (1<g<\infty )\]

\[
H_{-}(x,s)=-\frac{d^{2}}{dx^{2}}+V_{-}(x,s)=H_{+}(x,-s)\]

\[
V_{-}(x,s)=V_{+}(x,-s)\]

\[
H_{+}(x,s)=-\frac{d^{2}}{dx^{2}}+V_{+}(x,s)=H_{-}(x,-s)\]

\[
V_{+}(x,s)=V_{-}(x,-s),\]
 and (for example) \( H_{+}(x,s)\, y^{(-1/s)}=0 \) with \( H_{-}(x,s)\, y^{(-1/s)}\neq 0 \)
and \( y^{(-1/s)} \) is normalizable. In this regime, the ground state energy
of \( H_{+} \) is zero, and the other energy eigenvalues of both \( H_{-} \)
and \( H_{+} \) are paired. 

In other words, this duality exchanges the role of \( H_{+} \) and \( H_{-} \),
when one goes from the positive s regime to the negative s regime.
Roughly speaking, one could say that this duality exchanges the \char`\"{}boson\char`\"{}
spectrum and the \char`\"{}fermion\char`\"{} spectrum (see figure 1). In the
appendix, we work out two illustrative examples, one based on \( k(x)= \) constant
and the other one on the harmonic oscillator. Finally, thanks to a theorem of
Poincar\'e {[}7{]} one can prove that the solutions of the Sturm - Liouville equation
II.3 are analytic in \( s^{2} \), whatever \( k(x) \) is, if the boundary
conditions are independent of \( s \). In these situations, the ratio \( y'/y \)
will be \( s^{2} \) dependent and \( V_{-}(x,-s) \) will be equal to \( V_{+}(x,s) \).
That extends the cases of duality.\\
\\

\textbf{Conclusion}  \\

In this study we have shown that it is impossible to find a strong/weak duality
within a model of Supersymmetric Quantum Mechanics and we have developed a strategy
to find a class of models of SQM where the duality means a correspondence between
two regimes of the coupling constant.\\
\\

\textbf{Acknowledgments}\\

I would like to thank Guy Auberson (LPMT), Avinash Khare (Bhuba\-neswar Institute
of Physics, India) and Michel Talon (LPTHE, Paris VI) for their enlightening remarks and 
encouragements. I am also
grateful to members of the LPMT for many discussions, especially Christophe
Le Mouel, Gerard Mennessier and Gilbert Moultaka.\\
\\

\textbf{References}\\

(1) E. Witten, Nucl.Phys. B 185 (1981) 513 ; Nucl.Phys. B 202 (1982) 253. 

(2) A. Lahiri, P.K. Roy and B. Bagchi, Int.J.Mod.Phys. A Vol.5 No.8 (1990) 1383. 

(3) F.Cooper, A. Khare and U. Sukhatme, Phys.Rep. 251 (1995) 267. 

(4) S. Coleman, Phys.Rev.D 11(1975) 2088. 

(5) C. Wafa and E. Witten, Nucl.Phys. B 431 (1994) 3. 

(6) C. Hull and P. Townsend, Nucl.Phys. B 438 (1995) 109. 

(7) V. De Alfaro and T. Regge, Potential Scattering, North Holland Publishing Company (1965), page 9. \\
\\

\textbf{Appendix}\\
\renewcommand{\theequation}{A.\arabic{equation}}

\setcounter{equation}{0}

\underbar{Example 1} : \( k(\omega )=\omega ^{2}(\omega >0) \)\\

The master equation II.3 reads :

\[
y''-s^{2}\omega ^{2}y=0\, \, ,\]
 and an even solution is \( \cosh (s\omega x) \). One gets the potentials : 

\[
V_{-}(x,s)=\omega ^{2}\left( +s+(1-s)(\tanh (s\omega x))^{2}\right) \]
 and

\begin{equation}
\label{A.1}
V_{+}(x,s)=\omega ^{2}\left( -s+(1+s)(\tanh (s\omega x))^{2}\right) .
\end{equation}

It is easy to check that \( H_{-}(x,s) \) \( \left( \cosh (s\omega x)\right) ^{\left( \frac{1}{s}\right) }=0 \)
and that \( (\cosh (s\omega x))^{\left( \frac{1}{s}\right) } \) is normalizable
for negative \( s \). The duality is guaranteed since \( V_{-}(x,s)=V_{+}(x,-s) \).\\

The surprise in this example is that it is quite easy to find the full (discrete)
spectra of the hamiltonians because these potentials are in the S.I.P class
(Shape Invariant Potential) {[}2,3{]}. Indeed, potentials in A.1 can be derived
from the superpotential \( W=a\tanh (bx) \) when \( a=-\omega  \) and \( b=s\omega  \).
We refer to the reference {[}3{]} for details. One finds that the energy levels
are given by 

\[
E_{n}(s,\omega )=-ns\omega ^{2}(2+ns).\]
 Now, since \( s \) is negative and \( n \) is a non negative integer, and
because \( E_{n} \) must be an increasing function of \( n \), one gets that
\( n \) belongs to the interval \( \left[ 0,E(-1/s)\right]  \). {[}\( E(x) \)
\( \equiv  \) greatest integer less than or equal to \( x \){]}.

For example, if \( s=-1/4,\, H_{-} \) has 5 eigenvalues \( \left( 0,\, \omega ^{2}\frac{7}{16},\, \omega ^{2}\frac{12}{16},\, \omega ^{2}\frac{15}{16},\, \omega ^{2}\right)  \)
and \( H_{+} \) has 4 eigenvalues \( \left( \omega ^{2}\frac{7}{16},\, \omega ^{2}\frac{12}{16},\, \omega ^{2}\frac{15}{16},\, \omega ^{2}\right)  \).
See figure A1.\\

If one chooses \( k(x)=-\omega ^{2} \), one gets trigonometric functions instead
of hyperbolic functions. With this choice, \( x \) must be in an interval such
as \( \left] -\pi /2,\, +\pi /2\right[  \). \\

\underbar{Example 2} : \( k(x)=\omega ^{2}x^{2}(\omega >0) \)\\

The master equation II.3 reads:

\[
y''-s^{2}\omega ^{2}x^{2}y=0\, ,\]
 and an even solution is given in terms of a Bessel function :

\[
\sqrt{\left| x\right| }\, \textrm{Bessel}\, \textrm{I}\left[ -\frac{1}{4},(x^{2}\sqrt{\omega ^{2}s^{2})}/2\right] \]
 which has the following expansion (up to an overall numerical factor)

\[
1+\sum _{p=1}^{\infty }\left( \frac{s^{2p}\omega ^{2p}x^{4p}}{c_{p}}\right) \]
 where \( c_{p}=((3.4)(7.8)\cdots ((4p-1).(4p))) \).

Handling with care, the potentials read :

\[
V_{-}(x,s)=+s\omega x^{2}+(1-s\omega )x^{2}\left( \frac{\textrm{Bessel}\, \textrm{I}\left[ +\frac{3}{4},\, (s\omega x^{2})/2\right] }{\textrm{Bessel}\, \textrm{I}\left[ -\frac{1}{4},\, (s\omega x^{2})/2\right] }\right) ^{2}\]
 and

\begin{equation}
\label{A.2}
V_{+}(x,s)=-s\omega x^{2}+(1+s\omega )x^{2}\left( \frac{\textrm{Bessel}\, \textrm{I}\left[ +\frac{3}{4},\, (s\omega x^{2})/2\right] }{\textrm{Bessel}\, \textrm{I}\left[ -\frac{1}{4},\, (s\omega x^{2})/2\right] }\right) ^{2}
\end{equation}

Depending on \( s \), one obtains the same behaviour as in the example 1 and
the duality is got. The potentials are (probably) not S.I.P . In figure A2 are
drawn these potentials for \( s=-1/3 \) and \( \omega =1 \). For these values,
\( V_{-} \) shows two degenerate classically stable minima and one classically
unstable maximum. Tunneling effects in quantum mechanics make that the energy
of the ground state, which is of course \( 0 \) for supersymmetric reasons,
is at the level of the classical unstable extremum.\\

\textbf{Figure Captions} \\

\underbar{Figure 1} : typical discrete spectrum of a model in the negative s
regime on the left side and in the positive s regime on the right side
showing the \char`\"{}Bosonic\char`\"{} and \char`\"{}Fermionic\char`\"{} states
.\\

\underbar{Figure A1} : plots of \( V_{-} \)(thick) and \( V_{+} \) (thin)
and the energy eigenvalues (example A1 of the Appendix ; \( \omega =2,\, s=-1/4 \)).\\

\underbar{Figure A2} : plots of \( V_{-} \) (thick) and \( V_{+} \)(thin)
and the ground state energy (example A2 of the Appendix ; \( \omega =1,\, s=-1/3 \)).
\newpage
\vspace*{-1.cm}
$$\epsfig{file=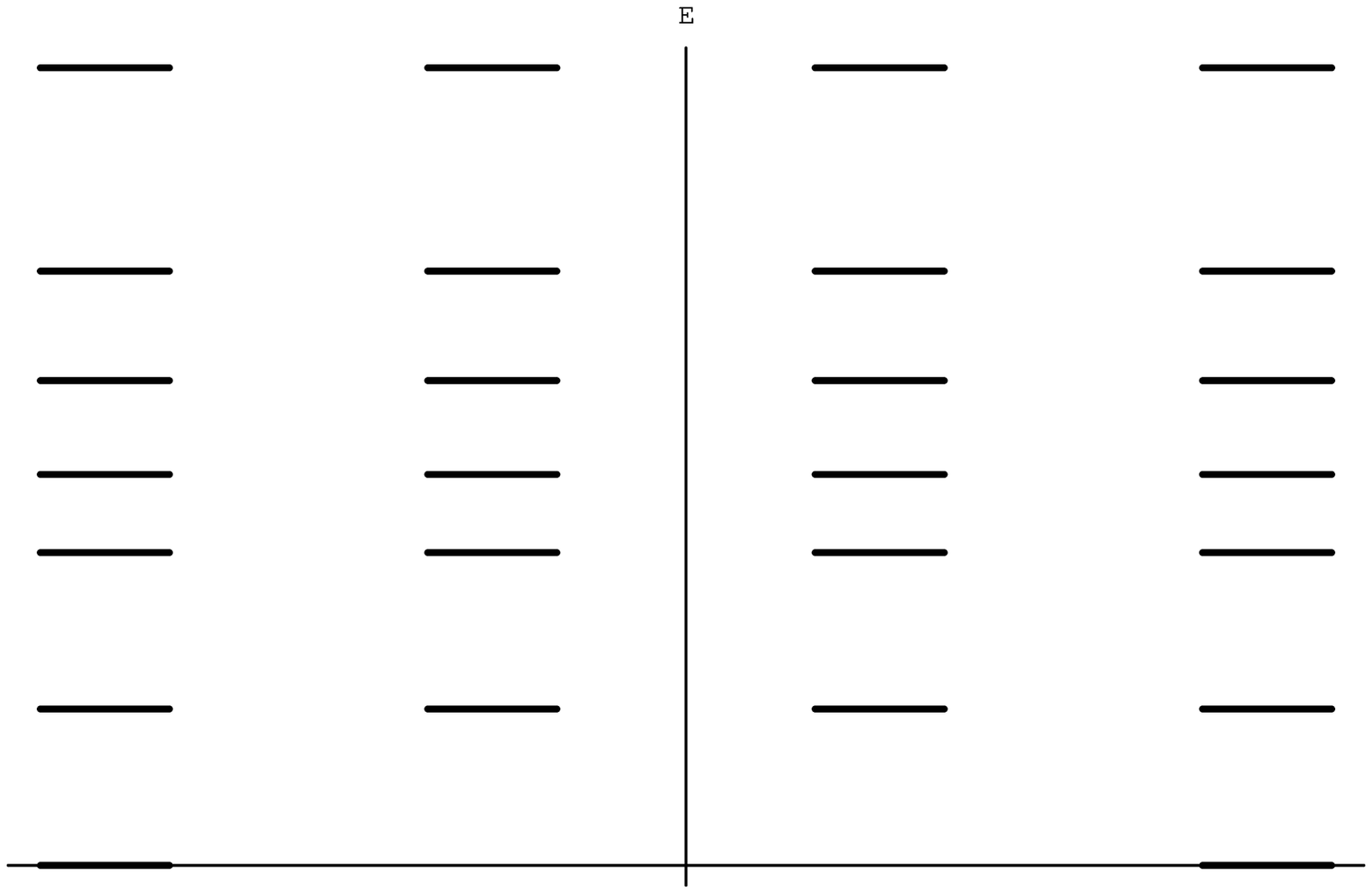,height=18.cm}$$
\null 
\vspace{-5.cm}\\

\hspace*{0.5cm} B  \hspace{3.2cm} F \hspace{3.2cm} B \hspace{3.2cm}
F\\[1cm]
\centerline{ Figure 1}
\newpage
\vspace*{-2.cm}
$$\epsfig{file=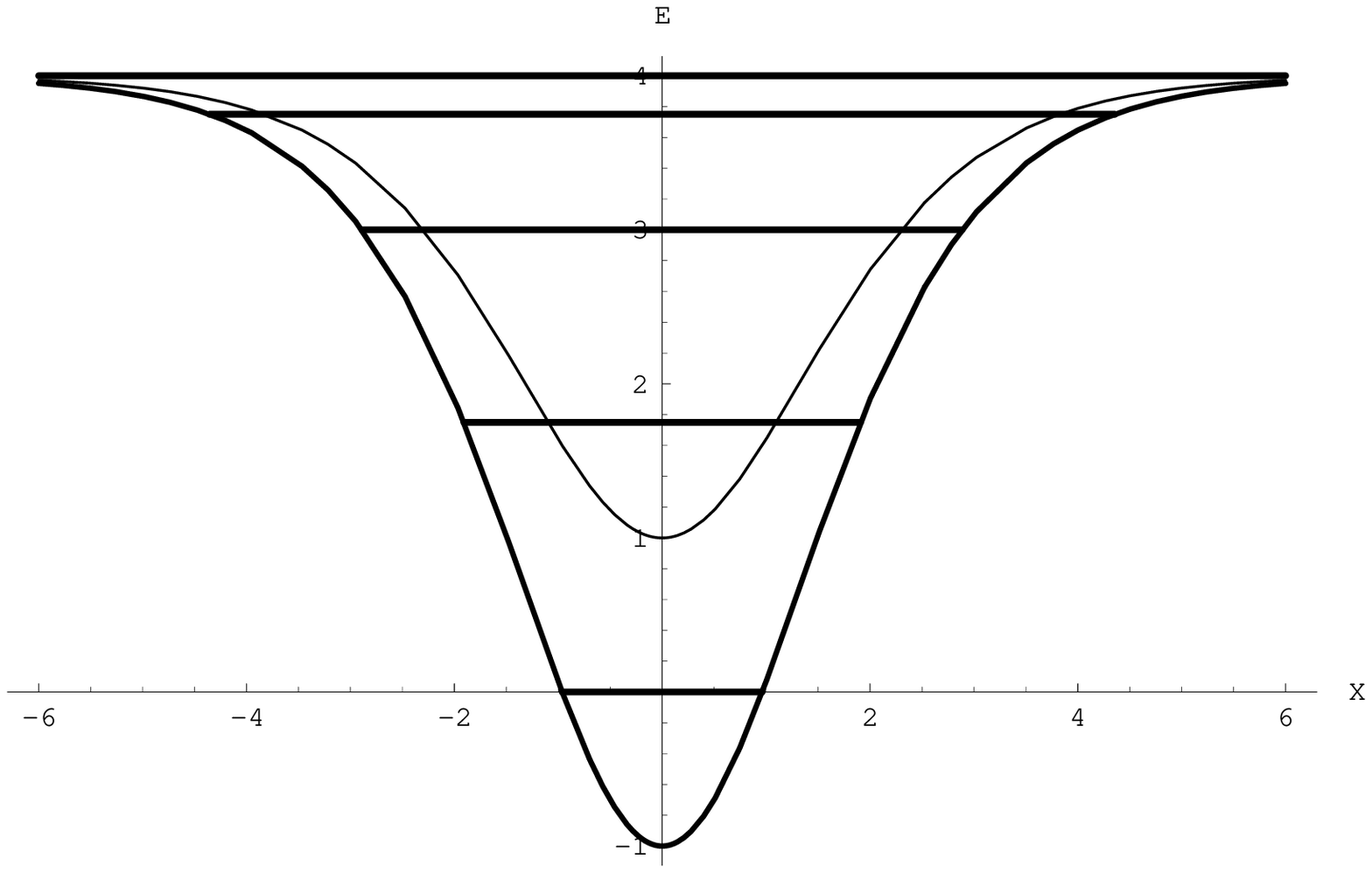,height=18.cm}$$
\null 
\vspace{-5.cm}\\
\centerline{ Figure A1}
\newpage
\vspace*{-2.cm}
$$\epsfig{file=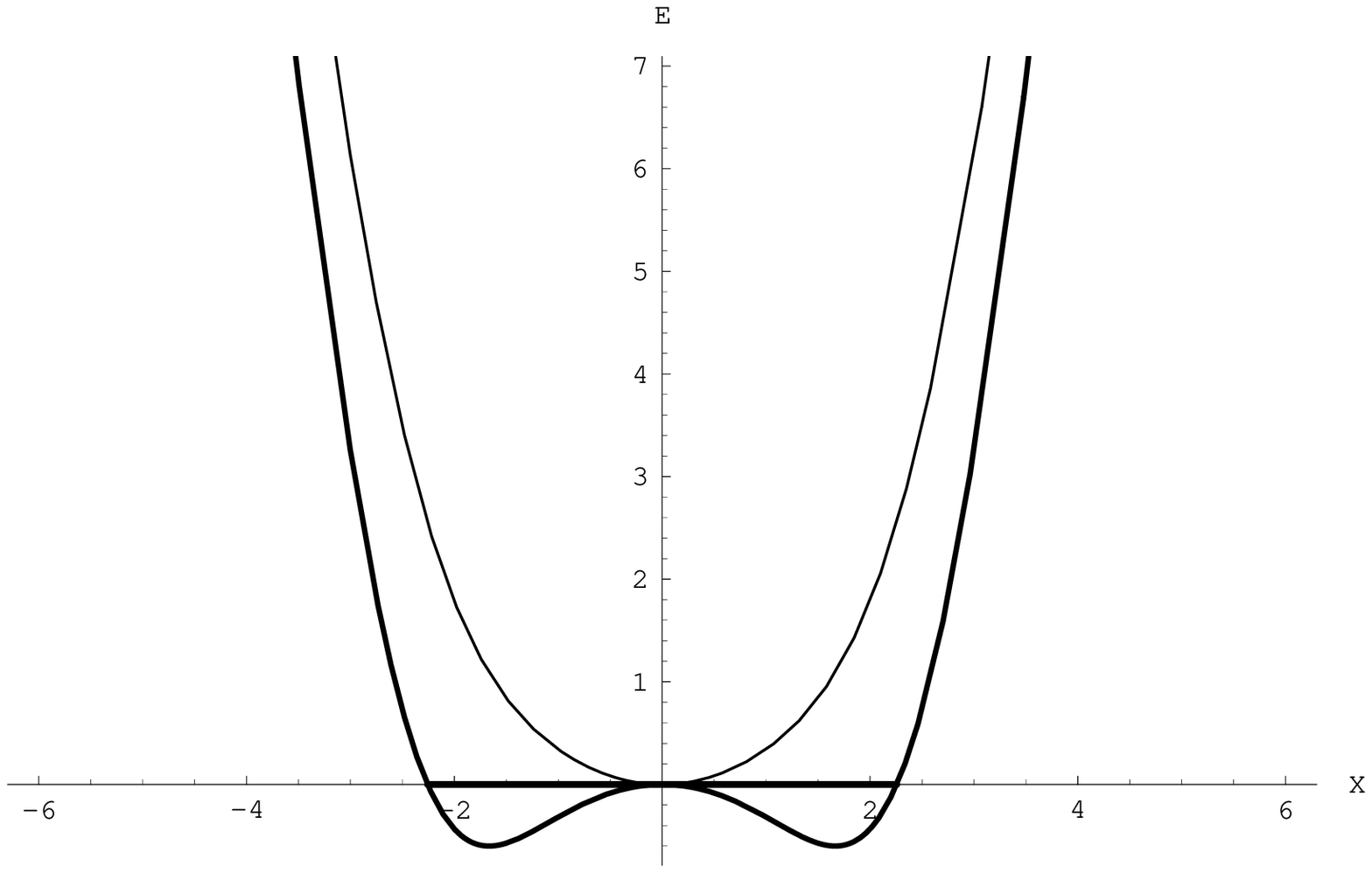,height=18.cm}$$
\null 
\vspace{-5.cm}\\
\centerline{ Figure A2}
\end{document}